\documentclass[11pt]{article}

\def\mydate{22 March 1999}

\newcommand{\beeq}{\begin{equation}}
\newcommand{\eneq}{\end{equation}}
\newcommand{\beqn}{\begin{eqnarray}}
\newcommand{\eeqn}{\end{eqnarray}}

\def\dd{\partial}
\def\la{\raise.16ex\hbox{$\langle$} \, }
\def\ra{\, \raise.16ex\hbox{$\rangle$} }
\def\ran{\raise.16ex\hbox{$\rangle$} }
\def\go{\rightarrow}

\def\next{{~~~,~~~}}
\def\onehalf{ \hbox{${1\over 2}$} }

\def\psibar{ \psi \kern-.65em\raise.6em\hbox{$-$}\lower.6em\hbox{} }
\def\Psibar{ \Psi \kern-.77em\raise.6em\hbox{$-$} }
\def\Phibar{ \Phi \kern-.77em\raise.6em\hbox{$-$} }

\def\vx{\vec x \,}
\def\L{ {\cal L} }
\def\H{ {\cal H} }
\def\ep{\epsilon}

\begin{document}

\thispagestyle{empty}

\baselineskip=12pt

{\small \noindent \mydate \hfill UMN-TH-1744/99}

{\small \noindent  \hfill EPHOU-99-003}

{\small \noindent  \hfill hep-th/9903193}

\baselineskip=15pt plus 1pt minus 1pt

\vskip 1.7cm

\begin{center}
{\LARGE \bf {The Hamilton-Jacobi Equations}}\\
\vspace{.3cm}
{\LARGE \bf {for Strings and p-Branes}}\\
\vspace{1.5cm}

{\large   Yutaka Hosotani}\\
\vspace{.1cm}
{\it School of Physics and Astronomy, University of Minnesota}\\ 
{\it  Minneapolis, MN 55455, U.S.A.}\\ 
\vspace{.5cm}
{\large   Ryuichi Nakayama}\\
\vspace{.1cm}
{\it Department of Physics, Faculty of Science, Hokkaido University}\\ 
{\it Sapporo  060-0810, Japan}\\ 
\end{center}

\vskip 1.cm

\begin{abstract}
Simple derivation of the Hamilton-Jacobi equation for bosonic strings and
p-branes is given.  The motion of classical strings and $p$-branes is
described by two  and $p+1$  local fields, respectively.  A variety of
local field equations which reduce to the Hamilton-Jacobi equation in the
classical limit are given.  They are essentially nonlinear, having no
linear term. 
\end{abstract}


\vskip 2.cm

\baselineskip=17pt plus 1pt minus 1pt


Dynamics of point particles can be described by the Hamilton-Jacobi (HJ)
equation.  It is a first-order partial differential equation,
given by $(\dd S)^2 = m^2$ for a free relativistic particle.  The
HJ equation can be viewed as the classical limit of 
quantum field equations.  The Klein-Gordon equation
$(\hbar^2 \dd^2 + m^2) \Phi =0$ or the Dirac equation $(i\hbar \gamma^\mu
\dd_\mu -m) \Psi=0$ reduces to the HJ equation with 
the ansatz $\Phi \sim \rho e^{iS/\hbar}$
or $\Psi \sim e^{iS/\hbar} \psi$.

The argument can be reversed for strings and p-branes.  We  attempt to
first find the classical HJ equations for strings and
p-branes, then use them as a guide to find quantum string and p-branes
equations.  In particular, establishing the Hamilton-Jacobi equation 
for classical superstrings would serve as a big step for an ultimate
formulation of quantum superstrings.

The Hamilton-Jacobi equation for strings has been discussed for many
years.\cite{Rinke1}-\cite{Hosotani2}  Rinke and Kastrup have given the HJ
equation for strings, though their derivation being rather invloved. For
point particles the HJ function $S(x)$ can be identified with the action of
trajectories ending at $x$.  Such identification is not applicable to
strings.  The purpose of this Letter is to give simple derivation of the
HJ equations for bosonic strings and p-branes, and to propose local field
equations which reduce to the HJ equations in the classical limit. We
believe that the simplicity and clarity of our derivation enhance the
usefulness
 of the HJ equation.

We start with the Nambu-Goto action for strings:
\beeq
I = - M^2 \int d\tau d\sigma  
\sqrt{ - \frac{1}{2} {v_{\mu\nu}}^2 } 
\label{NGaction}
\eneq
where $v^{\mu\nu} = \dd(x^\mu,x^\nu)/\dd(\tau,\sigma) $.
In terms of covariant momentum tensors
\beeq
p_{\mu\nu} = M^2 {v_{\mu\nu} \over \sqrt{ - v^2 /2} }
\next - {1\over 2} p_{\mu\nu} p^{\mu\nu} = M^4 
\label{momentum1}
\eneq
the equations of motions takes the form
$\dd ( p_{\mu\nu} , x^\nu )/\dd(\tau,\sigma) = 0 $.

The key step for the HJ formulation is to consider a family of solutions
to the equations such that they fill a $d$-dimensional domain in   
spacetime.  There are $d-2$ parameters, $\phi_a$'s, to specify these
solutions:  $x^\mu = x^\mu(\tau, \sigma; \phi_1, \cdots, \phi_{d-2})$.
They define a mapping from the parameter space 
$(\tau,\sigma, \phi_1, \cdots, \phi_{d-2})$ onto the spacetime domain
$( x^\mu )$.  In a region where the mapping is one-to-one, 
$p_{\mu\nu}(\tau, \sigma; \phi_a)$ can be viewed as a local field
$p_{\mu\nu}(x)$.  Then the equation of motion  is tranformed to
\beeq
p^{\mu\nu} \dd_\mu p_{\nu\lambda} =0 ~.
\label{eq1}
\eneq

Making use of the reparametrization invariance, one can choose
$(\tau, \sigma)$ such that $v_{\mu\nu}^2 = -2 $, independent
of $\tau$, $\sigma$ and $\phi_a$.  Then  
$M^{-2} p^{\mu\nu}d\tau \wedge d\sigma = dx^\mu \wedge dx^\nu$ is
the area element of the world sheet with fixed $\phi_a$'s.
Further  the two form 
$J(x) \equiv \onehalf p_{\mu\nu} dx^\mu \wedge dx^\nu = - M^2 d\tau \wedge
d\sigma$.  With the mapping between $\{ \tau,\sigma, \phi_a\}$ and 
$\{ x^\mu \}$,
$\tau$ and $\sigma$ are regarded as local fields.  Choosing
$S_1 = M \tau(x)$ and $S_2 = -M \sigma(x)$, $p_{\mu\nu}(x)$ is expressed
as 
\beeq
p_{\mu\nu} = \dd_\mu S_1 \dd_\nu S_2 - \dd_\nu S_1 \dd_\mu S_2 ~.
\label{momentum2}
\eneq

It is important to recognize that the equation of motion for strings 
is contained in the normalization condition.  Indeed with
(\ref{momentum2}),  Eq.\ (\ref{eq1}) becomes $p^{\mu\nu} \dd_\mu
p_{\nu\lambda}  = {1\over 2} p^{\mu\nu} \dd_{[\mu} p_{\nu\lambda ]} 
 - {1\over 4} \dd_\lambda (p^{\mu\nu} p_{\mu\nu})
=  - {1\over 4} \dd_\lambda (p^{\mu\nu} p_{\mu\nu}) =0$. Hence it is reduced
to the normailzation condition $p^{\mu\nu}p_{\mu\nu}=\,$constant, or to 
\beeq
(\dd S_1)^2 (\dd S_2)^2 - (\dd S_1 \dd S_2)^2 = - M^4 ~~.
\label{HJ1}
\eneq
Conversely, if the $S_1$ and $S_2$ satisfy the condition (\ref{HJ1}), then
$p_{\mu\nu}$ given by (\ref{momentum2}) satisfies the equation of motion.
From  $p_{\mu\nu}(x)$ a family of   the Nambu-Goto string solutions are 
reconstructed.   Eq.\ (\ref{HJ1}) is the Hamilton-Jacobi equation for
bosonic strings, first derived by Rinke.\cite{Rinke1}   The usage of a
family of solutions makes the proof significantly simple and transparent
compared with those of Rinke\cite{Rinke1} and of Kastrup\cite{Kastrup2}.  
The equivalence holds in any dimensions.  
Eqs.\ (\ref{momentum2}) and  (\ref{HJ1}) naturally
generalize  $p_\mu = \dd_\mu S$ and $(\dd S)^2 = m^2$ for point
particles.  

Kastrup, Nambu and Rinke proposed a relation $p_{\mu\nu} =  \dd_\mu A_\nu
- \dd_\nu A_\mu$ a long time ago.\cite{Kastrup1} We see $A_\mu = S_1
\dd_\mu S_2$ up to a gauge tranformation.  The vector potential
description is redundant.   It was shown by one of the authors  that
string motion can be expressed in terms of $(d-2)$ local scalar
fields,\cite{Hosotani1}  which is unnecessarily  general and 
redundant.

The above argument applies to the point particle case as well.  The
equation of motion is
$\dot p^\mu =0$ where $p^\mu = m\dot x^\mu/\sqrt{\dot x^2}$.   A family of
solutions are parametrized by $d-1$ $\phi_a$'s; $x^\mu = x^\mu(\tau;
\phi_1, \cdots, \phi_{d-1})$.  With this mapping the equation is converted
to  $p^\mu \dd_\mu p_\lambda =0$.  With the parametrization $\dot x^2=1$,
$p_\mu dx^\mu = m d\tau$ so that
the covarinat momentum can be expressed
as $p_\mu = \dd_\mu S$.   The equation fo motion is reduced to  $p^\mu
\dd_\mu p_\lambda  = p^\mu \dd_\lambda p_\mu = {1\over 2} \dd_\lambda p^2
=0$, i.e.\
$(\dd S)^2 = m^2$.  If one considers a family of trajectories starting
at one point $x^\mu_{(0)}$ at $\tau=0$, $\phi_a$'s are just momenta
at $x^\mu_{(0)}$.  With this choice $S(x)$ is the action evaluated at $x$.
In the case of strings the  meaning of 
$S_1(x)$ and $S_2(x)$ is yet to be found.

The generalization to p-branes is straightforward.  The action for p-branes
in the Nambu-Goto form is
\beeq
I = - M^{p+1} \int d\tau d^p \sigma 
\left[ ~ {(-)^p v^2\over (p+1)!}   ~ \right]^{1/2} 
\label{action2}
\eneq
where $v^{\mu_1 \cdots \mu_{p+1}} =\dd(x^{\mu_1}, \cdots, x^{\mu_{p+1}})
/\dd (\tau, \sigma_1, \cdots, \sigma_p)$. The covariant momentum
tensors are given by $p^{\mu_1 \cdots \mu_{p+1}} = 
M^{p+1} v^{\mu_1 \cdots \mu_{p+1}} ~ [ (-)^p v^2/(p+1)!]^{-1/2}$. The
equation of motion is given by
$\dd( p_{\mu_1 \cdots \mu_{p+1}} , x^{\mu_1}, \cdots, x^{\mu_{p}} )
/ \dd (\tau, \sigma_1, \cdots, \sigma_p )  = 0$. 
Again by considering a family of solutions the  equation is converted
to 
\beeq
p^{\mu_1 \cdots \mu_{p+1}} \dd_{\mu_1} p_{\mu_2 \cdots \mu_{p+1} \nu}
 =0 ~. 
\label{eq2}
\eneq
With the parametrization $v^2 = (-1)^p (p+1)!$, 
$p_{{\mu_1} \cdots {\mu{p+1}}} dx^{\mu_1} \wedge \cdots \wedge
dx^{\mu_{p+1}} \propto d\tau \wedge d\sigma_1 \wedge \cdots \wedge
d\sigma_p$.   The covariant momentum tensor field is represented in terms
of $(p+1)$ local scalar fields:
\beeq
p_{\mu_1 \cdots \mu_{p+1}} 
= {\dd( S_1, \cdots, S_{p+1}) \over 
     \dd ( x^{\mu_1}, \cdots, x^{\mu_{p+1}}) } ~.
\label{HJ2}
\eneq
In this representation 
$p^{\mu_1 \cdots \mu_{p+1}} \dd_{\mu_1} p_{\mu_2 \cdots \mu_{p+1} \nu}
=(-1)^p \dd_\nu ( p_{\mu_1 \cdots \mu_{p+1}}  p^{\mu_1 \cdots \mu_{p+1}})
/2(p+1)$ so that the equation of motion becomes equivalent to
the Hamilton-Jacobi equation  given by 
\beeq
{1\over (p+1)!} \left\{ {\dd( S_1, \cdots, S_{p+1}) \over 
     \dd ( x^{\mu_1}, \cdots, x^{\mu_{p+1}}) }  \right\}^2 
= (-1)^p M^{2(p+1)} ~~.
\label{HJ3}
\eneq
The equation for membranes ($p=2$) has been discussed by 
Aurilia et al.\cite{Aurilia}
 
Quantum equations which reduce to the Hamilton-Jacobi equation in the
classical limit $\hbar \go 0$ can be easily found.   There are 
several options.   In the case of strings ($p$-branes) one may start with a
functional of lines ($p$-dimensional surfaces).  This approach leads to
the string ($p$-brane) field theory.  In this Letter we propose an
alternative approach.  We look for quantum equations in the form of  local
field equations, which turn out essentially nonlinear with no ``free''
part.  Being local field equations, they are expected to describe
only a part of dynamics of quantum strings.  Yet it is  interesting and
noteworthy that they have connection to classical string dynamics through
the HJ equation.

The first candidate is given by 
\beeq
\L_1 = - {\hbar^4\over 2}  ~ \Sigma_{\mu\nu}^\dagger  \Sigma^{\mu\nu}
 - M^4 \Phi_1^\dagger \Phi_1 \Phi_2^\dagger \Phi_2 
\next \Sigma_{\mu\nu} = {\dd (\Phi_1, \Phi_2) \over \dd (x^\mu, x^\nu)}
\label{lagrangian1}
\eneq
where $\Phi_a$'s are complex scalar fields.
The Euler equations reduce to the HJ equation (\ref{HJ1}) in the
$\hbar \go 0$ limit with the ansatz $\Phi_a = \rho_a e^{i S_a/\hbar}$. 
The Lagrangian (\ref{lagrangian1}) contains no bilinear terms.
The Hamiltonian density is positive semi-definite: 
$\H_1  = \hbar^4 (\Sigma_{0k}^\dagger \Sigma_{0k} 
+ {1\over 2} \Sigma_{jk}^\dagger \Sigma_{jk} ) 
+ M^4 \Phi_1^\dagger \Phi_1 \Phi_2^\dagger \Phi_2$.  However, expressed in
terms of conjugate momenta 
$\Pi_a^\dagger =  \hbar^4 \epsilon_{ab} \Sigma_{0k}^\dagger \dd_k \Phi_b$,
it appears singular when $\nabla \Phi_1 \propto \nabla \Phi_2$:
\beeq
\H_1 = {1\over \hbar^4} 
{\Pi_a^\dagger \dd_j \Phi_a^\dagger \dd_j \Phi_b \Pi_b
\over |\nabla \Phi_1 |^2 |\nabla \Phi_2|^2 
   - |\nabla\Phi_1^\dagger  \nabla\Phi_2|^2 }
+ {\hbar^4 \over 2} ~ \Sigma_{jk}^\dagger \Sigma_{jk} 
  + M^4 \Phi_1^\dagger \Phi_1 \Phi_2^\dagger \Phi_2  ~.
\label{hamiltonian1}
\eneq
This is an essentially nonlinear system.  
Although the canonical quantization can be carried out, the full 
consistency is yet to be examined.

The second candidate is obtained by generalizing Dirac's approach.
We prepare two multi-component fields $\Phi$ and $\Psi$ and write
\beeq
\L_2 = \hbar^2 \Psibar_a \Phibar_b ~ \Gamma^{\mu\nu}_{ab,cd} ~
   \dd_\mu \Phi_d \dd_\nu \Psi_c - M^2 \Psibar_a\Psi_a  \Phibar_b\Phi_b ~.
\label{lagrangian2}
\eneq
With the ansatz $\Psi = e^{iS_1/\hbar} \psi$ and 
$\Phi = e^{iS_2/\hbar} \phi$, the Euler equations reduce in the $\hbar \go
0$ limit to
\beeq
\Gamma^{\mu\nu}_{ab,cd} \dd_\mu S_1 \dd_\nu S_2
 + M^2 \delta_{ac}\delta_{bd} =0 ~~,
\label{eq4}
\eneq
or in short $\Gamma^{\mu\nu} \dd_\mu S_1 \dd_\nu S_2 + M^2 =0$.  This
equation becomes identical to the HJ equation (\ref{HJ1}), provided
\beeq
\{ \Gamma^{\mu\nu} , \Gamma^{\rho\sigma} \}_{ab,cd}
 = -2 (g^{\mu\rho} g^{\nu\sigma} - g^{\mu\sigma} g^{\nu\rho} )
\delta_{ac} \delta_{bd} ~~.
\label{algebra1}
\eneq

$\Gamma^{\mu\nu}$'s act on the two fields.  If $\Psi$ and $\Phi$
have $n_1$ and $n_2$ components, respectively, then $\Gamma^{\mu\nu}$'s are
$n_1 n_2$ dimensional matrices.  To find representations of the algebra
(\ref{algebra1}), first consider the same algebra obeyed by $n$-dimensional
matrices $\gamma^{\mu\nu}$:
$\{ \gamma^{\mu\nu} , \gamma^{\rho\sigma} \} 
 = -2 (g^{\mu\rho} g^{\nu\sigma} - g^{\mu\sigma} g^{\nu\rho} )$.  
Since $\{ \gamma^{\mu\nu} \}$ defines  $d(d-1)/2$ dimensional Clifford
algebra in $d$ dimensions, $n= 2^{[d(d-1)/4]}$ in the minimal
representation.  This algebra for $\gamma^{\mu\nu}$'s 
has been previously investigated in the  context of string and p-brane
field equations in refs.\ \cite{Hosotani3} and \cite{Ho1}.

Suppose that $\Phi$ is $m$-component scalar field and $\Psi_a$
is in a ``spinor'' representation of the $\gamma^{\mu\nu}$ algebra.
In other words we write
\beeq
\Gamma^{\mu\nu}_{ab,cd} = \gamma^{\mu\nu}_{ac} v_{bd} ~~,
\label{rep1}
\eneq
which satisfies (\ref{algebra1}) provided $v_{ab} v_{bc} = \delta_{ac}$.
The simplest choice is to consider a single component field $\Phi$
($m=1$),  which leads to
\beeq
\L_3 = \hbar^2 \Psibar \gamma^{\mu\nu} \dd_\nu \Psi \cdot
\Phi^\dagger \dd_\mu \Phi - M^2 \Psibar \Psi \cdot \Phi^\dagger \Phi ~.
\label{lagrangian3}
\eneq

The Lorentz transformation properties of $\Psi$  are clarified 
by constructing the corresponding generators.  Under a Lorentz 
transformation $\delta x^\mu = {\ep^\mu}_\nu x^\nu$, 
$\delta \Psi_a  = -{i\over 2} \ep_{\mu\nu} s^{\mu\nu}_{ab}
\Psi_b$.  As is easily confirmed, 
$s^{\mu\nu} = - {i\over 4} [\gamma^{\mu\alpha} , {\gamma^\nu}_\alpha ]$
satisfies the desired Lorentz algebra.  The Dirac conjugate is given by
$\Psibar = \Psi^\dagger \omega$  where
$\omega = \omega^\dagger = (i ~{\rm or}~1) \prod_{j=1}^{d-1}\gamma^{0j}$
for $d=4n$ or $4n+2$, respectively.  Further
$-i [s^{\rho\sigma}, \gamma^{\mu\nu}]
= g^{\mu\rho} \gamma^{\nu\sigma} + g^{\nu\sigma} \gamma^{\mu\rho}
 - g^{\mu\sigma} \gamma^{\nu\rho} - g^{\nu\rho} \gamma^{\mu\sigma}$ and
$\omega^{-1} s_{\mu\nu}^\dagger \omega = s_{\mu\nu}$.  These guarantee
the invariance of  the Lagrangian (\ref{lagrangian3})  under proper Lorentz
transformations.  In $d=4$, for instance, $\Psi$ has 8 components, which
consists of two vectors as can be seen from $s^{\mu\nu}$.\cite{Hosotani3}

The discrete symmetry properties are subtle, however.  Under parity $P$,
$(x^0, \vx) \go (x'{}^0, \vx')= (x^0, -\vx) $ and
$\Psi(x) \go \Psi'(x') = U_P \Psi(x)$.  In order for the Lagrangian 
(\ref{lagrangian3}) to be invariant under $P$, we need
(i)  $U_P^\dagger \omega U_P = \omega$ and 
(ii)  $U_P^\dagger (\gamma^{0k}, \gamma^{jk}) U_P = 
(-\gamma^{0k}, \gamma^{jk})$. These two are incompatible in even 
dimensions.  The Lagrangian (\ref{lagrangian3}) is not invariant
under $P$.

Another choice is to consider $\Psi$ and $\Phi$ in the same 
$n$-plet representation of the $\gamma^{\mu\nu}$ algebra.  Depending on the
dimensionarity, however, one needs two copies of either $\Psi$ or $\Phi$.
In $d=4$ ($mod$ 4), $\bar\gamma = \prod \gamma^{\mu\nu}$ (up to a phase)
satisfies $\{ \bar\gamma, \gamma^{\mu\nu} \} =0$,
$\bar\gamma^\dagger=\bar\gamma$, and $\bar\gamma^2 =1$. We observe
\beeq
\Gamma^{\mu\nu}_{ab,cd}
=  {1\over \sqrt{2}} \left( \gamma^{\mu\nu}_{ac} \delta_{bd}
 + \bar\gamma_{ac} \gamma^{\mu\nu}_{bd} \right) ~, 
\label{algebra2}
\eneq
or in short $\Gamma^{\mu\nu} = (\gamma^{\mu\nu} \otimes I +
\bar\gamma \otimes \gamma^{\mu\nu})/\sqrt{2}$,  satisfies the
algebra (\ref{algebra1}).  The corresponding Lagrangian is
\beeq
\L_4 = {\hbar^2\over \sqrt{2}} 
\Big\{ \Psibar \gamma^{\mu\nu} \dd_\nu \Psi \cdot \Phibar \dd_\mu \Phi
- \Phibar \gamma^{\mu\nu} \dd_\nu\Phi\cdot \Psibar \bar\gamma \dd_\mu \Psi 
\Big\} - M^2 \Psibar\Psi \cdot\Phibar\Phi ~.
\label{lagrangian4}
\eneq
The Lorentz generators are given by
$S^{\mu\nu} = s^{\mu\nu} \otimes I + I \otimes s^{\mu\nu}$.  
As $[\bar\gamma , s^{\mu\nu}]=0$, the  Lagrangian $\L_4$ is Lorentz
invariant.    The invariance of $\L_4$ under $P$ is achieved if there
exists $U_P$ such that  (i)  $U_P^\dagger \omega U_P = +\omega$ or
$-\omega$,  and  (ii)  $U_P^\dagger (\gamma^{0k}, \gamma^{jk}, \bar\gamma)
U_P =  (-\gamma^{0k}, \gamma^{jk}, \bar\gamma)$.   The conditions for
$\gamma^{0k}$ and $\bar\gamma$ are incompatible.  $\L_4$ in the minimal
representation is not  invariant under $P$ in $d=4$ ($mod$ 4).
In $d=2$ ($mod$ 4) the dimension of $\Psi$ must be doubled to have
$\bar\gamma$.  With the doubling the Lagrangian can be made $P$ invariant
in even dimensions. Write $\Gamma^{\mu\nu} = (\hat\gamma^{\mu\nu} \otimes I
+\bar\gamma \otimes \gamma^{\mu\nu})/\sqrt{2}$ in place of 
(\ref{algebra2}).  The Dirac conjugate of $\Psi$ is written as $\Psibar =
\Psi^\dagger \hat\omega$.  Under $P$, $\Psi \go \hat U_P \Psi$.   The
appropriate choice is : $\hat\gamma^{\mu\nu} = \gamma^{\mu\nu} \otimes
\tau_1$,
$\bar\gamma = I \otimes \tau_3$, $\hat U_P = i \omega \otimes \tau_2$.
$\hat\omega$ can be either $\omega \otimes 1$ or $\omega \otimes \tau_2$.

In this paper we have given simple derivation of the
Hamilton-Jacobi  equations for bosonic strings and p-branes, Eqs.\
(\ref{HJ1}) and (\ref{HJ3}).   We  have then explored local
field equations which reduce to the  HJ equations in the classical limit. 
There are a variety of possibilities. In all cases the equations are
essentially nonlinear, which makes the  quantization formidable.  We have
not known whether the systems proposed are well defined at the quantum
level.  The critical dimensionality inherent in string theories must arise
at the quantum level.   The
$\Gamma^{\mu\nu}$ algebra (\ref{algebra1}) is naturally associated with
strings.  Yet its implementation in field theories in (\ref{lagrangian3})
and (\ref{lagrangian4}) requires more elaboration.

As remarked before, the local field equations proposed above are expected 
to describe at best a part of string dynamics.  Ultimate string
field equations should be formulated in loop space with their projection
yielding the local field equations proposed here.  The extension of the 
current formalism to superstrings is yet to be achieved.  We believe that
the  Hamilton-Jacobi equations should serve as a helpful guide in searching
such ultimate superstring field equations.

\newpage

\noindent {\bf Acknowledgements}

\vskip .3cm

This work was supported in part  by the Ministry of Education, Science,
Sports and Culture, Japan under the Grant-in-Aid for International
Science Research (Grant No.\ 10044043), and  by the U.S.\ Department of
Energy under contracts DE-FG02-94ER-40823.  R.N.\ and Y.H.\ would like to
thank the School of Physics and Astronomy, University of Minnesota, and
the Department of Physics, Hokkaido University, respectively, for 
their hospitality where a part of the work was done.

\def\jnl#1#2#3#4{{#1}{\bf #2} (#4) #3}
\def\Zphys{{\em Z.\ Phys.} }
\def\jssc{{\em J.\ Solid State Chem.\ }}
\def\jpsJ{{\em J.\ Phys.\ Soc.\ Japan }}
\def\ptps{{\em Prog.\ Theoret.\ Phys.\ Suppl.\ }}

\def\JMP{{\em J. Math.\ Phys.} }
\def\NPB{{\em Nucl.\ Phys.} B}
\def\NP{{\em Nucl.\ Phys.} }
\def\PLB{{\em Phys.\ Lett.} B}
\def\PL{{\em Phys.\ Lett.} }
\def\PRL{\em Phys.\ Rev.\ Lett. }
\def\PRB{{\em Phys.\ Rev.} B}
\def\PRD{{\em Phys.\ Rev.} D}
\def\PRe{{\em Phys.\ Rep.} }
\def\AP{{\em Ann.\ Phys.\ (N.Y.)} }
\def\RMP{{\em Rev.\ Mod.\ Phys.} }
\def\ZPC{{\em Z.\ Phys.} C}
\def\SCI{\em Science}
\def\CMP{\em Comm.\ Math.\ Phys. }
\def\MPLA{{\em Mod.\ Phys.\ Lett.} A}
\def\IJMPB{{\em Int.\ J.\ Mod.\ Phys.} B}
\def\PR{{\em Phys.\ Rev.} }
\def\cmp{{\em Com.\ Math.\ Phys.}}
\def\JPA{{\em J.\  Phys.} A}
\def\CQG{\em Class.\ Quant.\ Grav. }

\vskip .7cm

\leftline{\bf References}  

\renewenvironment{thebibliography}[1]
        {\begin{list}{[$\,$\arabic{enumi}$\,$]}  
        {\usecounter{enumi}\setlength{\parsep}{0pt}
         \setlength{\itemsep}{0pt}  \renewcommand{\baselinestretch}{1.2}
         \settowidth
        {\labelwidth}{#1 ~ ~}\sloppy}}{\end{list}}

\end{document}